\documentclass[aps,prl,twocolumn,showpacs,amsmath,amssymb,amsfonts,nofootinbib,long]{revtex4}
\usepackage{epsfig,latexsym,bm}

\newcommand\keV{\mbox{keV}}
\newcommand\MeV{\mbox{MeV}}
\newcommand\GeV{\mbox{GeV}}
\newcommand\TeV{\mbox{TeV}}
\newcommand\Pl{\mbox{\scriptsize Pl}}

\begin{document}

\title{Supersymmetric Q-balls: A Numerical Study}

\author{L. Campanelli$^{1,2}$}
\email{campanelli@fe.infn.it}
\author{M. Ruggieri$^{3,4}$}
\email{marco.ruggieri@ba.infn.it}

\affiliation{$^1$Dipartimento di Fisica, Universit\`{a} di Ferrara, I-44100 Ferrara, Italy}
\affiliation{$^2$INFN - Sezione di Ferrara, I-44100 Ferrara, Italy}
\affiliation{$^3$Dipartimento di Fisica, Universit\`{a} di Bari, I-70126 Bari, Italy}
\affiliation{$^4$INFN - Sezione di Bari, I-70126 Bari, Italy}

\date{October, 2007}


\begin{abstract}
We study numerically a class of non-topological solitons, the
Q-balls, arising in supersymmetric extension of the Standard Model
with low-energy, gauge-mediated symmetry breaking.
Taking into account the exact form of the supersymmetric potential
giving rise to Q-balls, we find that there is a lower limit on the
value of the charge $Q$ in order to make them classically stable:
$Q \gtrsim 5 \times 10^2 Q_{\rm cr}$, where $Q_{\rm cr}$ is
constant depending on the parameters defining the potential and
can be in the range $1 \lesssim Q_{\rm cr} \lesssim 10^{8 \div
16}$.
If $Q$ is the baryon number, stability with respect to the decay
into protons requires $Q \gtrsim 10^{17} Q_{\rm cr}$, while if the
gravitino mass is greater then $m_{3/2} \gtrsim 61 \MeV$, no
stable gauge-mediation supersymmetric Q-balls exist.
Finally, we find that energy and radius of Q-balls can be
parameterized as $E \sim \xi_E Q^{3/4}$ and $R \sim \xi_R
Q^{1/4}$, where $\xi_E$ and $\xi_R$ are slowly varying functions
of the charge.
\end{abstract}


\maketitle



\section{I. Introduction}

Cosmological and astrophysical observations indicate, almost
undoubtedly, that our Universe is pervaded by a not-yet-known
pressureless component, called ``Dark Matter''~\cite{DarkMatter}.
Its energy density amount to about the $30\%$ of the entire energy
density of the universe.
\\
Promising non-baryonic, dark-matter candidates have been proposed
since the discover of this mysterious component, such as axion or
neutralino~\cite{DarkMatter}.

Recently enough~\cite{Kusenko-Shaposhnikov}, a particular class of
non-topological solitons arising in supersymmetric extensions of
the Standard Model, know as supersymmetric Q-balls~\cite{Dvali},
have been proposed as possible solution to the dark matter
problem.
\\
In particular, Q-balls admitted in supersymmetric models with
low-energy, gauge-mediated symmetry breaking~\cite{SUSY}, are a
plausible candidate for baryonic dark matter~\cite{KusenkoTalk}.

Q-balls~\cite{Coleman} are lumps of matter, precisely a coherent
state of a complex scalar field, carrying a conserved global
charge. In the context of supersymmetric extensions of the
Standard Model, the charge $Q$ is some combination of baryon and
lepton numbers, while the the scalar field is a gauge-singlet
combination of squarks and sleptons corresponding to some flat
direction of the supersymmetric potential~\cite{Dvali}. In this
class of models, supersymmetry is spontaneously broken at the
scale $\Lambda_{\rm DSB} \sim 10^7 \GeV$~\cite{SUSY}; as a result,
an effective potential for the flat directions arises~\cite{de
Gouvea} which, in turn, admits Q-balls as the non-perturbative
ground state of the theory.

A deep investigation of general properties~\cite{Q-balls-1} and
astrophysical implications~\cite{Q-balls-2} of Q-balls has been
carried out in the last two decades.
\\
Up to now, however, the main properties of gauge-mediation
supersymmetric Q-balls have been analyzed using an approximate
expression of the potential giving rise to Q-balls.

The aim of this paper is to study such a kind of Q-balls taking
into account the exact form of the supersymmetric scalar potential
quoted in Ref.~\cite{de Gouvea}. We find that the expressions for
energy and radius of Q-balls, which fully characterize their
properties from a cosmological and astrophysical viewpoint, can
differ from the approximate case of about an order of magnitude.


\section{II. Q-balls: General Properties}

In this Section, we briefly review the Q-ball solution of a scalar
theory with a global $U(1)$ symmetry~\cite{Coleman}. We consider a
charged scalar field $\phi$ whose lagrangian density is given by
\begin{equation}
\label{eq:Lagr1} \mathcal{L} = (\partial_\mu
\varphi^*)(\partial^\mu \varphi) - U(|\varphi|^2).
\end{equation}
In the next Section, we will identify $\phi$ with one of the flat
directions in supersymmetric extensions of the standard model, and
will specify the form of the potential $U(|\varphi|)$ which is not
relevant in the present discussion. For the moment, we simply
require its invariance under a global $U(1)$ symmetry. The
corresponding conserved Noether charge $q$ is normalized as
\begin{equation}
\label{eq:charge} q = \frac{1}{2i} \int \! d^3x \left(\varphi^*
\dot{\varphi} - \varphi \dot{\varphi}^* \right) \!,
\end{equation}
where a dot indicates a derivative with respect to time.
(Throughout this paper, we follow the conventions of
Ref.~\cite{Kusenko}). For a given field configuration
$\varphi(t,{\bm r})$, the total energy is given by
\begin{equation}
\label{eq:energy} E = \int \! d^3x \left[\frac{1}{2} \,
|\dot{\varphi}|^2 + \frac{1}{2} \, |{\nabla \varphi}|^2 +
U(|\varphi|^2) \right] \!.
\end{equation}
We are interested to solutions of the field equations that
correspond to a fixed value of the charge, namely $Q$, in
Eq.~\eqref{eq:charge}. This is properly achieved by the
introduction of the Lagrange multiplier $\omega$ associated to
$q$, and by the requirement that the physical configuration makes
the functional
\begin{equation}
\label{functional1} \mathcal{E}_\omega \equiv E + \omega \left[Q -
\frac{1}{2i} \int \! d^3x \left(\varphi^* \dot{\varphi} - \varphi
\dot{\varphi}^* \right) \right]
\end{equation}
stationary with respect to independent variations of $\varphi$ and
$\omega$~\cite{Kusenko}.
The requirement of time-independence of the total energy $E$
implies the choice~\cite{Coleman,Kusenko}
\begin{equation}
\varphi(t,{\bm r}) = e^{i\omega t} \phi({\bm r}),
\end{equation}
with $\phi({\bm r})$ real. Consequently, the functional
$\mathcal{E}_\omega$ reduces to
\begin{equation}
\label{eq:energyW2} \mathcal{E}_\omega = \int \! d^3x \left[
\frac{1}{2} \, |{\nabla\phi}|^2 + U(\phi) - \frac{1}{2} \,
\omega^2 \phi^2 \right] + \omega Q.
\end{equation}
The physical solutions have to satisfy the constraints
\begin{equation}
\label{eq:PhysCond} \frac{\delta \mathcal{E}_\omega}{\delta\phi} =
0, \;\;\;\; \frac{\delta \mathcal{E}_\omega}{\delta\omega} = 0.
\end{equation}
The first constraint leads to the equation of motion of the field
$\phi({\bm r})$,
\begin{equation}
\label{eq:eqMot3} \frac{d^2 \phi}{dr^2} + \frac{2}{r}\frac{d
\phi}{d r} + \omega^2 \phi = \frac{\partial U}{\partial \phi} \,.
\end{equation}
where, for simplicity, we assumed isotropy: $\phi = \phi(r)$ with
$r \equiv |{\bm r}|$.
\\
The second constraint is equivalent to the requirement that the
charge corresponding to the solution of the equation of motion is
equal to $Q$.
\\
A Q-ball is defined as the solution $\phi(r)$ of
Eq.~\eqref{eq:eqMot3} satisfying, at fixed charge $Q$, the
boundary conditions $\phi(r \rightarrow \infty) = 0$ and
$d\phi/dr(r=0) = 0$~\cite{Coleman}.

In the next Section, we will analyze Q-ball configurations arising
in a supersymmetric model where supersymmetry is broken via
low-energy gauge mediation~\cite{SUSY}.


\section{III. Supersymmetric Q-balls}\label{Sec:SUSYq}

We consider a supersymmetric model in which supersymmetry is
broken by low-energy gauge mediation~\cite{SUSY}. In this kind of
model the coupling of the massive vector-like messenger fields to
the gauge multiplets, with coupling constant $g\sim 10^{-2}$,
leads to the breaking of supersymmetry~\cite{SUSY}. The coupling
itself gives rise to an effective potential for the flat direction
$\phi$ whose lowest order (two-loop) contribution has been
calculated in Ref.~\cite{de Gouvea}:
\begin{equation}
\label{potential} U(z) = \Lambda \! \int_0^1 \!\! dx \,
\frac{z^{-2} - x(1-x) + x(1-x)\ln[x(1-x)z^2]}{[z^{-2}-x(1-x)]^2}
\, .
\end{equation}
Here, $z \equiv \phi/M$ and $M \equiv M_S/(2g)$, with $M_S$ the
messenger mass scale. The value of the mass parameter
$\Lambda^{1/4}$ is constrained as (see, e.g.,
Ref.~\cite{Kasuya511}):
$10^3 \GeV \lesssim \Lambda^{1/4} \lesssim (g^{1/2}/4\pi)
\sqrt{m_{3/2} M_{\Pl}}$,
where $M_{\Pl} \sim 2.4 \times 10^{18} \GeV$ is the reduced Planck
mass and $m_{3/2}$, the gravitino mass, is in the range $100 \,
\keV \lesssim m_{3/2} \lesssim 1\GeV$~\cite{de Gouvea,Kasuya511}.

The asymptotic expressions of $U(z)$, for small and large $z$
are~\cite{de Gouvea}:
\begin{equation}
\label{potentialapprox} \frac{U(z)}{\Lambda} \simeq
    \left\{ \begin{array}{ll}
        z^2, &  \;\; \mbox{if} \;\; z \ll 1, \\
        (\ln z^2)^2 - 2 \ln z^2 + \frac{\pi^2}{3} \, , &  \;\; \mbox{if} \;\; z \gg 1,
    \end{array}
    \right.
\end{equation}
where $m \equiv \sqrt{2 \Lambda}/M$ is the soft breaking mass and
is of order $1\TeV$~\cite{Kasuya511}. In Fig.~1, we plot the
potential $U(z)$ with its asymptotic expansions
\eqref{potentialapprox}.


\begin{figure}[t]
\begin{center}
\includegraphics[clip,width=0.4\textwidth]{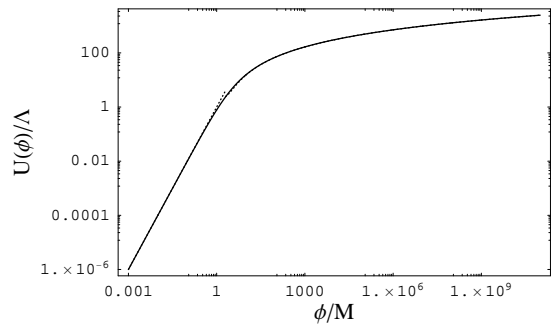}
\caption{The potential $U(\phi)$. Dotted lines refer to the
asymptotic expansions for small ($\phi \ll M$) and large values
($\phi \gg M$) of the field $\phi$ [see
Eq.~(\ref{potentialapprox})].}
\end{center}
\end{figure}


Defining the critical charge $Q_{\rm cr} \equiv \Lambda/m^4$
(whose meaning will be clear in the following), the constraint on
$\Lambda$ can be translated to a constraint on $Q_{\rm cr}$,
namely:
\begin{equation}
\label{Qcr} \left( \frac{\TeV}{m} \right)^{\!\!4} \lesssim \;
Q_{\rm cr} \, \lesssim 10^8 \left( \frac{g}{10^{-2}} \right)^{\!2}
\left( \frac{\TeV}{m} \right)^{\!\!4} \left( \frac{m_{3/2}}{100 \,
\keV} \right)^{\!2} \!\!.
\end{equation}
A widely used approximation consists in replacing the full
potential $U(z)$ with its asymptotic expansions
(\ref{potentialapprox}) in which a plateau plays the role of the
logarithmic rise for large values of $z$.
\\
Within this approximation, it has been shown that the potential
$U(z)$ allows Q-balls solutions as the non perturbative ground
state of the model~\cite{Dvali,Kusenko-Loveridge}. Such states are
known as supersymmetric Q-balls. In particular, for large charges,
$Q \gg Q_{\rm cr}$, one can deduce analytically the most important
characteristics of Q-ball solitons. In more detail, the Q-ball
profile is given by~\cite{Dvali,Kusenko-Loveridge}:
$\phi(r) \simeq \phi_0 \sin (\omega r)/(\omega r)$ for $r \leq R$,
and zero for $r \geq R$, where $R \equiv \pi/\omega$ is the radius
of the Q-ball, and $\phi_0 \equiv \phi(0)$. Moreover, one
has~\cite{Dvali,Kusenko-Loveridge}
\begin{eqnarray}
\label{omegaL} && \frac{\omega}{m} \simeq \sqrt{2} \pi
\left(\frac{Q}{Q_{\rm cr}}\right)^{\! -1/4} \! ,
\\
\label{EnergyL} && \frac{E}{m Q_{\rm cr}} \simeq
\frac{4\sqrt{2}\pi}{3} \left(\frac{Q}{Q_{\rm cr}}\right)^{\! 3/4}
\! ,
\\
\label{raggioL} && \frac{R}{m^{-1}} \simeq \frac{1}{\sqrt{2}}
\left(\frac{Q}{Q_{\rm cr}}\right)^{\! 1/4} \! ,
\\
\label{phi0L} && \frac{\phi_0}{M} \simeq \frac{1}{\sqrt{2}}
\left(\frac{Q}{Q_{\rm cr}}\right)^{\! 1/4} \! .
\end{eqnarray}
Although the previous (simplified) analysis reveals the major
properties of supersymmetric Q-balls, which are widely used in the
literature in a cosmological and astrophysical context, we wish to
study them by taking into account the full potential
\eqref{potential}. Since the form of the potential is involved, we
need to solve the problem numerically.

The computational procedure has been depicted in the previous
Section. From a numerical viewpoint, it is simpler to fix $\omega$
rather than the total charge $q$. Then, once $\omega$ is fixed, we
solve the equation of motion~\eqref{eq:eqMot3} with the condition
$d\phi/dr(r=0) = 0$. We look for the value of $\phi_0$ such that
the Q-ball solution exists; once this is achieved we insert
$\phi(r)$ in Eqs.~\eqref{eq:charge} and \eqref{eq:energy},
obtaining the values of the charge and energy. Finally, we define
the ``radius'' of the Q-ball, $R$, such that $\phi(R)/\phi_0 =
0.1$.


\begin{figure}[t]
\begin{center}
\includegraphics[clip,width=0.4\textwidth]{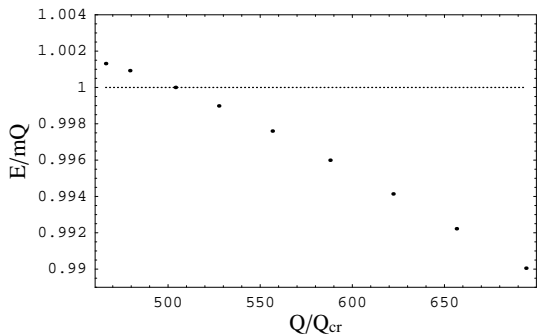}
\caption{The Q-ball energy $E$ normalized to the energy of $Q$
quanta of the field $\phi$ (that constitute the perturbative
spectrum of the theory) as a function of the charge.}
\end{center}
\end{figure}


In Fig.~2, we plot the quantity $E/mQ$ as a function of the
charge. If the energy $E$ of the Q-ball at fixed charge $Q$ is
less then $mQ$, the soliton decays into $Q$ quanta of the field
(the perturbative spectrum of the theory), each of them with mass
$m$. Instead, if $E < mQ$ the Q-ball is said to be classically
stable, and then represents the ground state of the theory.
\\
Numerically, we find classical stability, $E/mQ < 1$, for $Q >
Q_{\rm min}$, with $Q_{\rm min} \simeq 504 Q_{\rm cr}$.


\begin{figure}[t]
\begin{center}
\includegraphics[clip,width=0.4\textwidth]{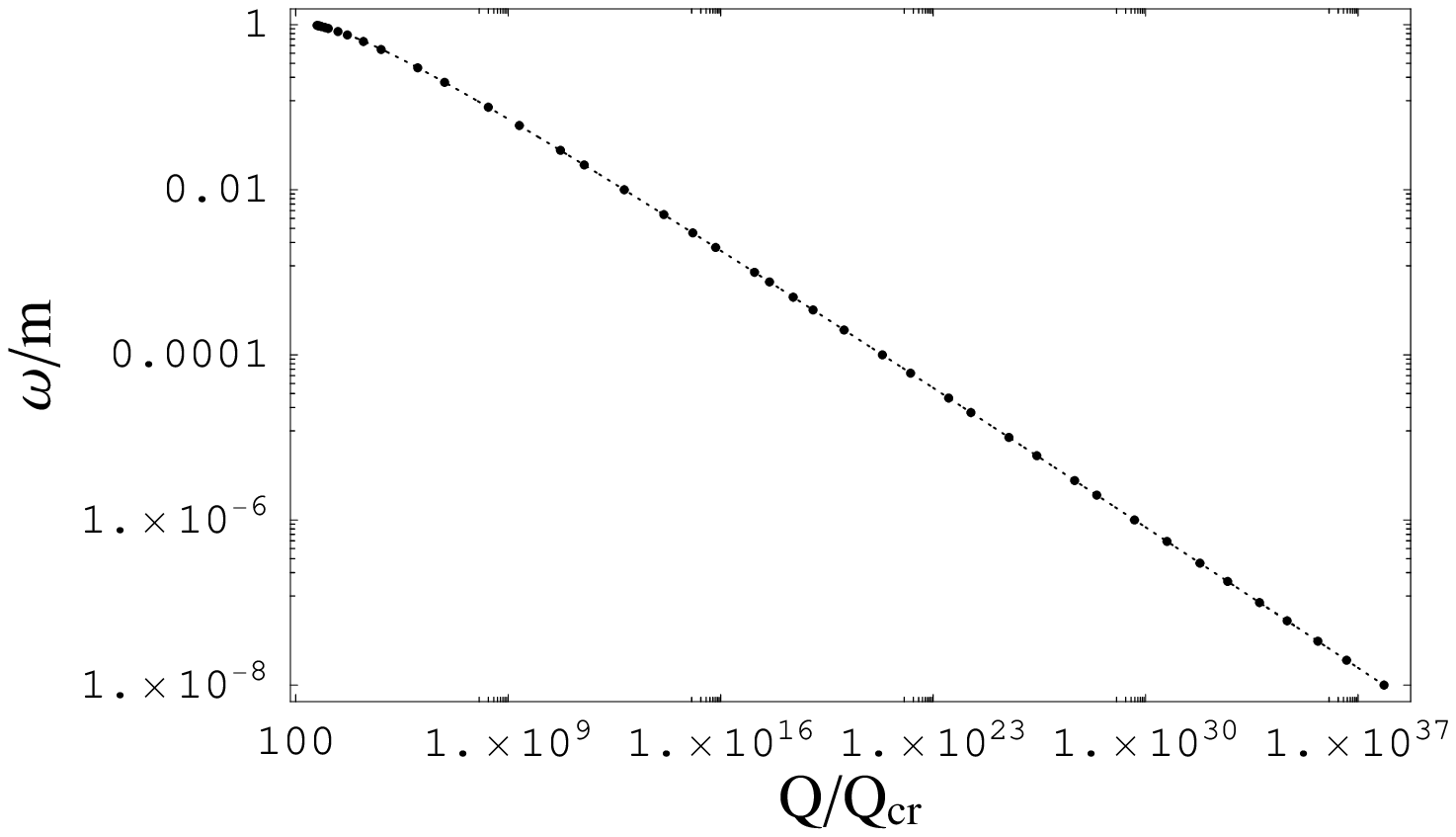}
\\
\includegraphics[clip,width=0.4\textwidth]{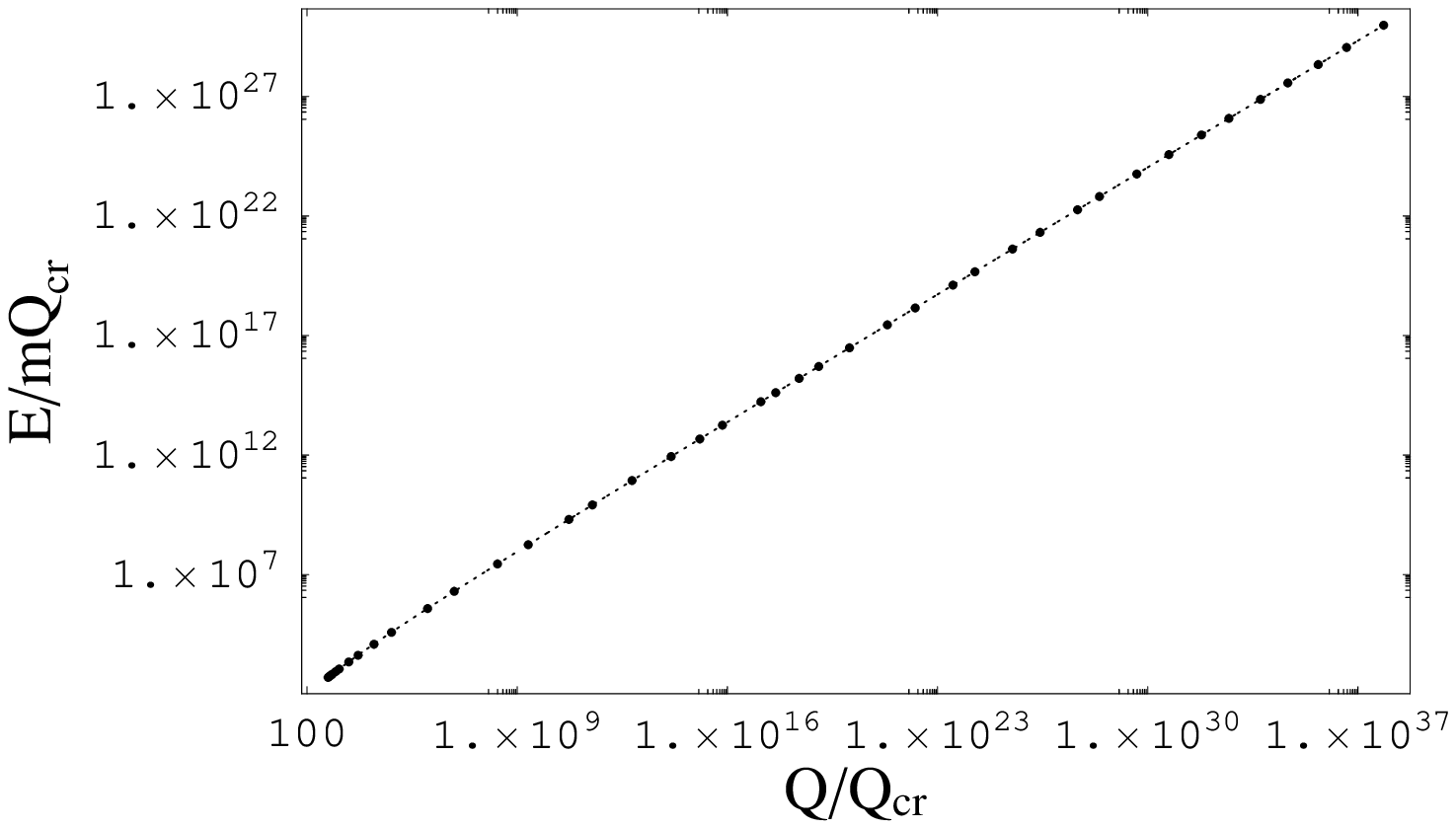}
\\
\includegraphics[clip,width=0.4\textwidth]{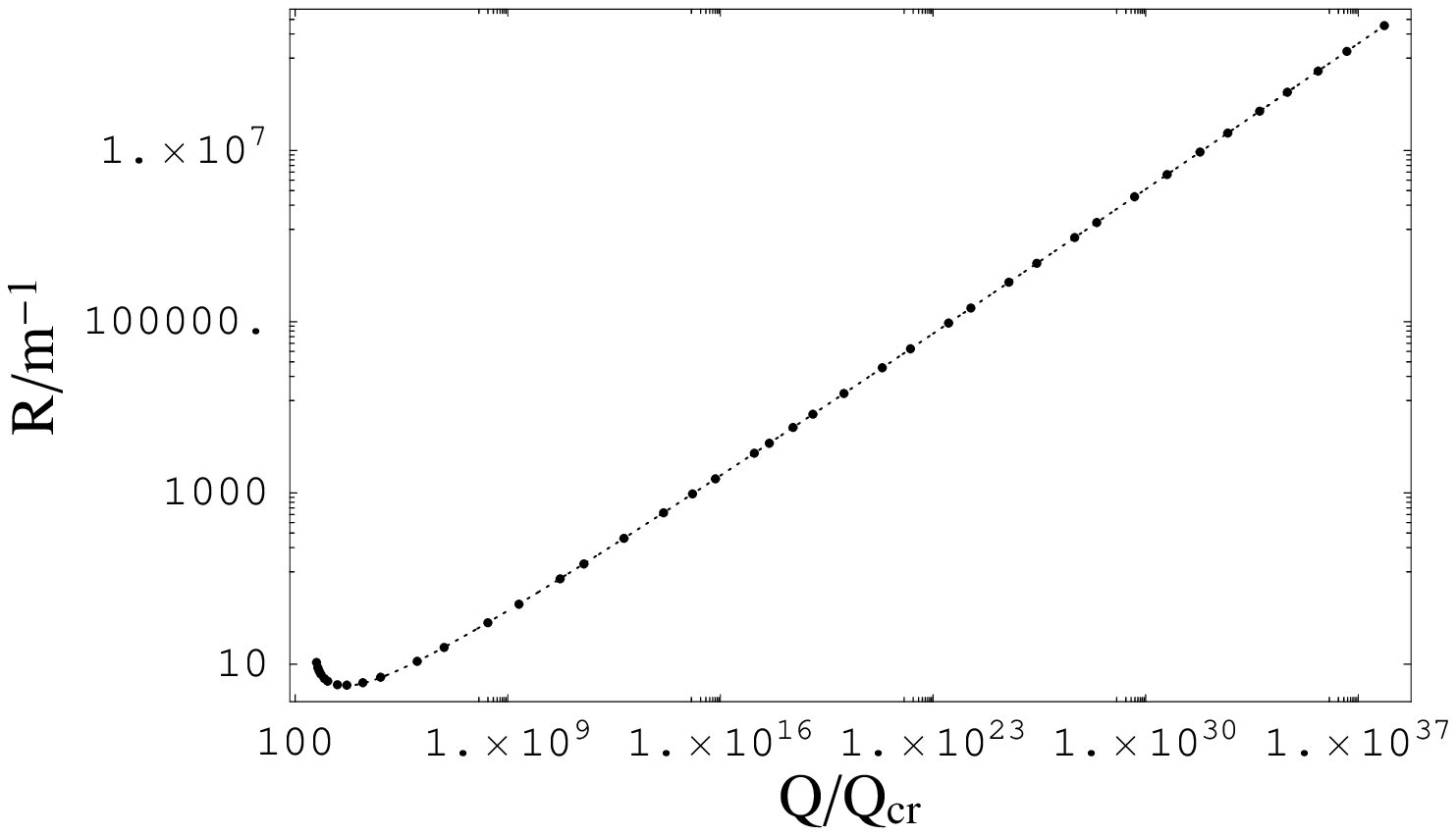}
\\
\includegraphics[clip,width=0.4\textwidth]{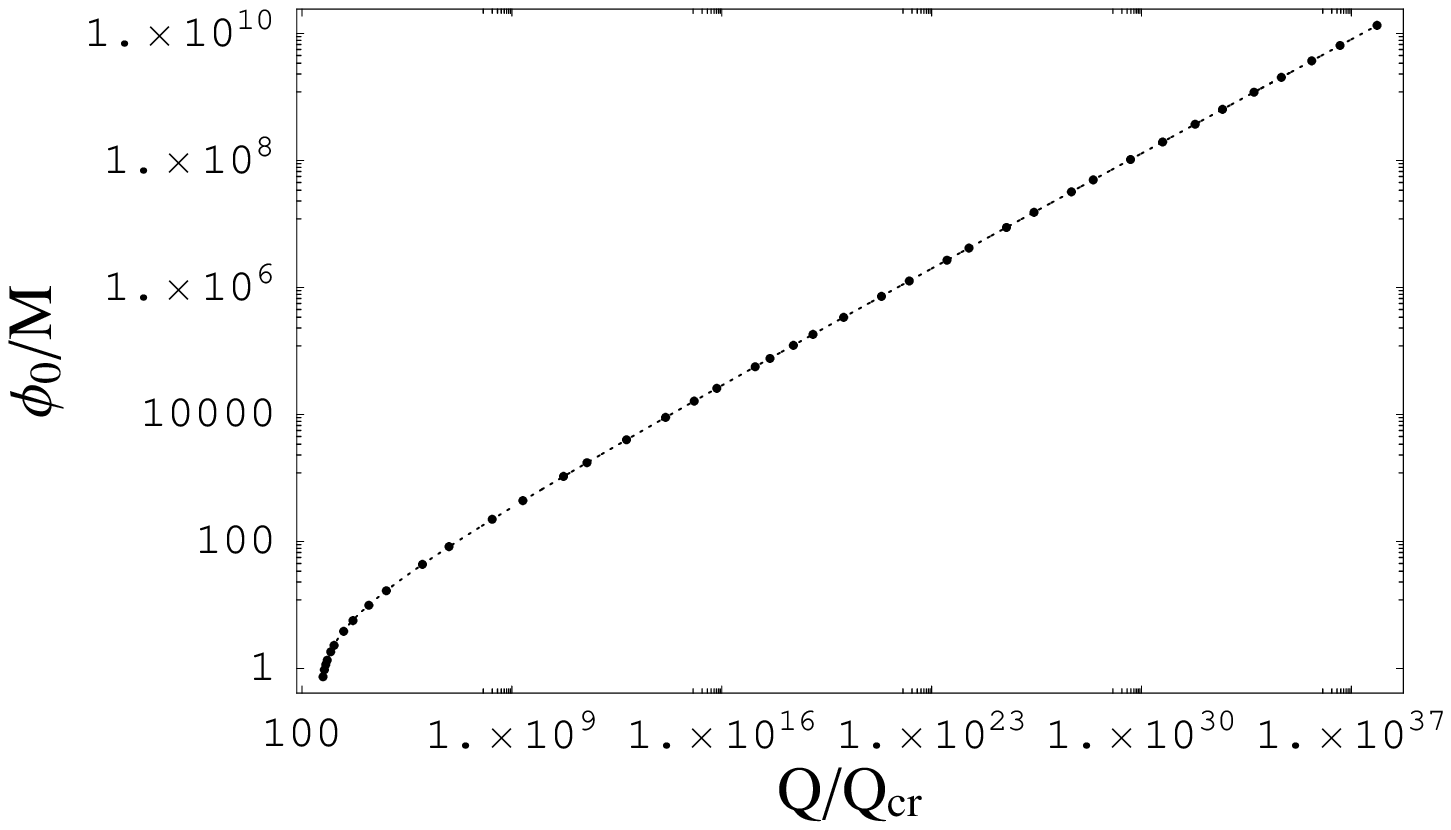}
\caption{From top to down: the parameter $\omega$, the energy $E$,
the radius $R$, and $\phi_0$ as a function of the charge $Q$
normalized to the critical charge $Q_{\rm cr}$. Dotted lines refer
to the approximate solutions (\ref{omega})-(\ref{phi0}),
respectively [see also Eq.~(\ref{xi}) and Table I].}
\end{center}
\end{figure}


In Fig.~3, we plot $\omega$, the energy, the radius $R$, and
$\phi_0$, as a function of the charge in the interval $Q \in
[Q_{\rm min}, 7.2 \times 10^{37} Q_{\rm cr}]$.

If $Q$ is the baryon number, stability with respect to the decay
into protons requires the energy of a Q-ball to be less then $E <
m_p Q$, where $m_p \simeq 1 \GeV$ is the proton mass. We find
numerically that this is attained for charges larger then $Q
\gtrsim 4.1 \times 10^{17} Q_{\rm cr}$, where we assumed $m = 1
\TeV$.

We now wish to compare our numerical results to those obtained in
the flat-potential approximation. Inspired by
Eqs.~\eqref{omegaL}-\eqref{phi0L}, we write the quantities
characterizing the Q-ball solution in the following way:
\begin{eqnarray}
\label{omega} && \frac{\omega}{m} = \xi_\omega \! \left(
\log_{10}\frac{Q}{Q_{\rm cr}} \right) \left(\frac{Q}{Q_{\rm
cr}}\right)^{\! -1/4} \! ,
\\
\label{energy} && \frac{E}{m Q_{\rm cr}} = \xi_E \! \left(
\log_{10}\frac{Q}{Q_{\rm cr}} \right) \left(\frac{Q}{Q_{\rm
cr}}\right)^{\! 3/4} \! ,
\\
\label{raggio} && \frac{R}{m^{-1}} = \xi_R \! \left(
\log_{10}\frac{Q}{Q_{\rm cr}} \right) \left(\frac{Q}{Q_{\rm
cr}}\right)^{\! 1/4} \! ,
\\
\label{phi0} && \frac{\phi_0}{M} = \xi_\phi \! \left(
\log_{10}\frac{Q}{Q_{\rm cr}} \right) \left(\frac{Q}{Q_{\rm
cr}}\right)^{\! 1/4} \! .
\end{eqnarray}
The functions $\xi$'s (which depend only logarithmically on the
charge $Q$) parameterize the deviation from the simple power-laws
(\ref{omegaL})-(\ref{phi0L}), and are shown in Fig.~4. We fit
their numerical values by the power-function
\begin{equation}
\label{xi} \xi(x) = (a + b x^p)^q.
\end{equation}
In Table I, we report the values of the coefficients $a$, $b$,
$p$, and $q$ found by least-squaring the numerical data. We also
show the maximum percentage error of the functions $\xi$'s with
respect to their numerical values. In particular, $\varepsilon_1$
and $\varepsilon_2$ refer to the maximum percentage errors in the
ranges $Q \in [Q_{\rm min}, 7.2 \times 10^{37} Q_{\rm cr}]$ and $Q
\in [10^{17} Q_{\rm cr}, 7.2 \times 10^{37} Q_{\rm cr}]$,
respectively.

In is worth noting that, in the flat-potential approximation, the
functions $\xi$'s are constants whose values differ from the
numerical results of about an order of magnitude in the limit of
large charges (say $Q \gg 10^{17} Q_{\rm cr}$).


\begin{figure}[t]
\begin{center}
\includegraphics[clip,width=0.4\textwidth]{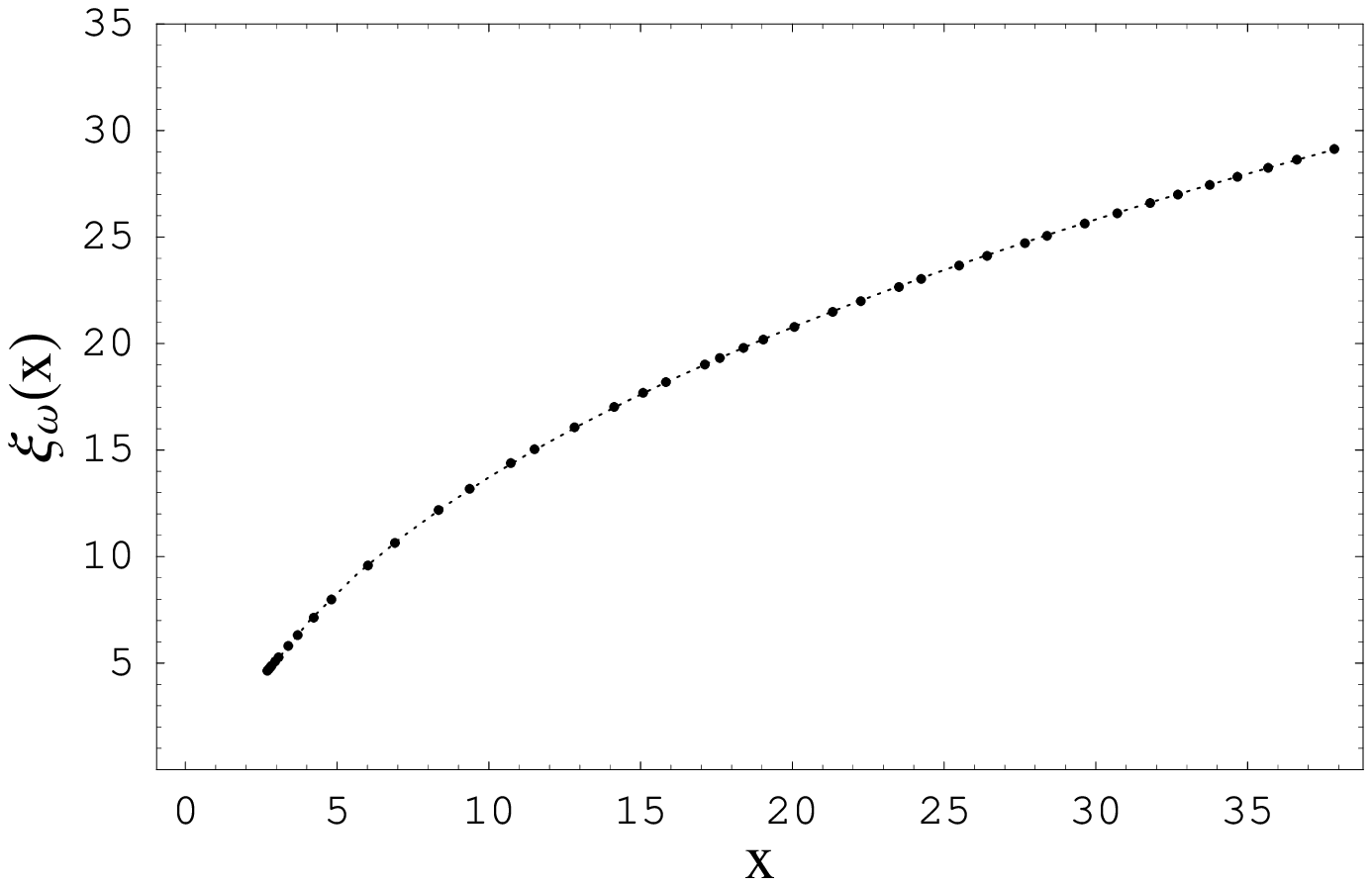}
\\
\includegraphics[clip,width=0.4\textwidth]{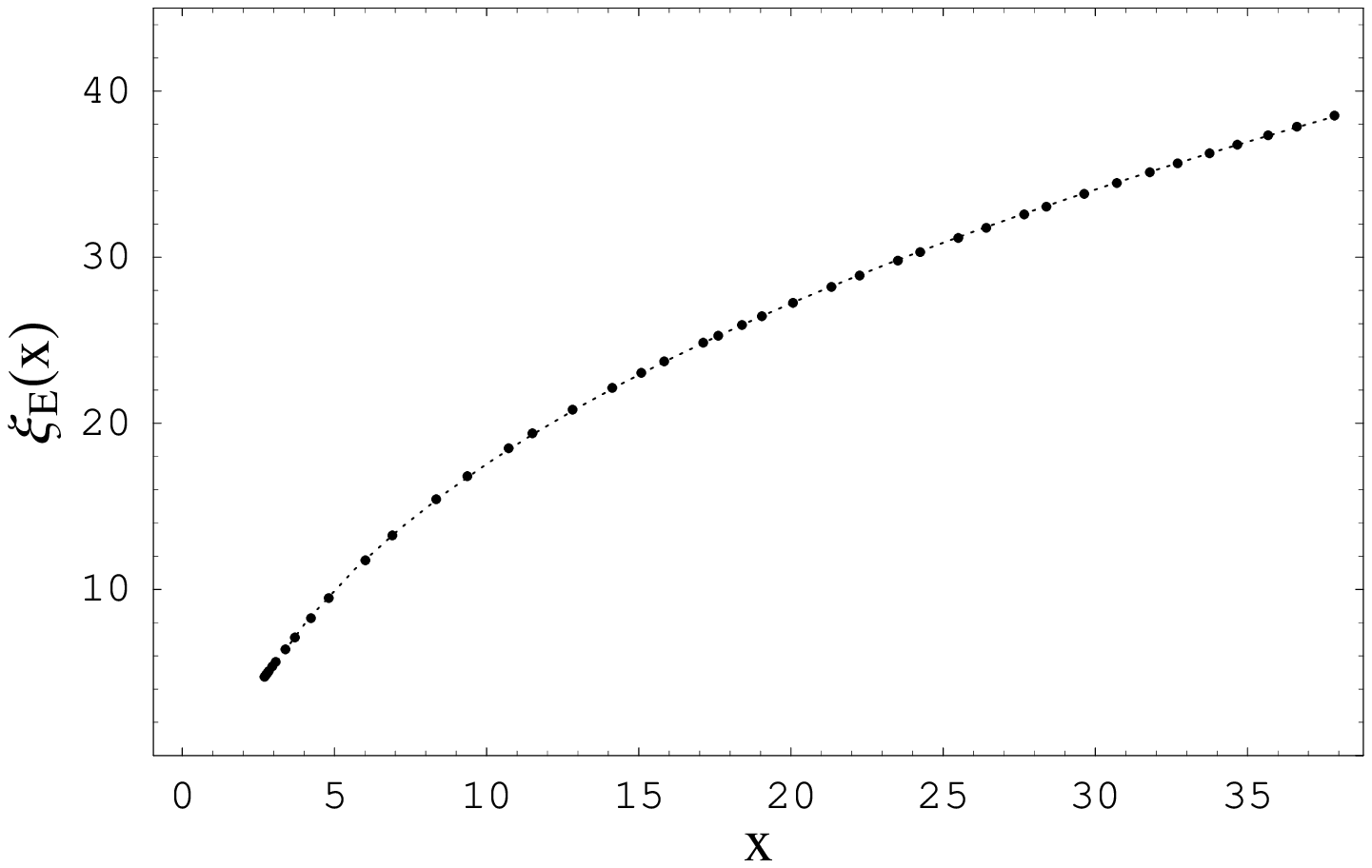}
\\
\includegraphics[clip,width=0.4\textwidth]{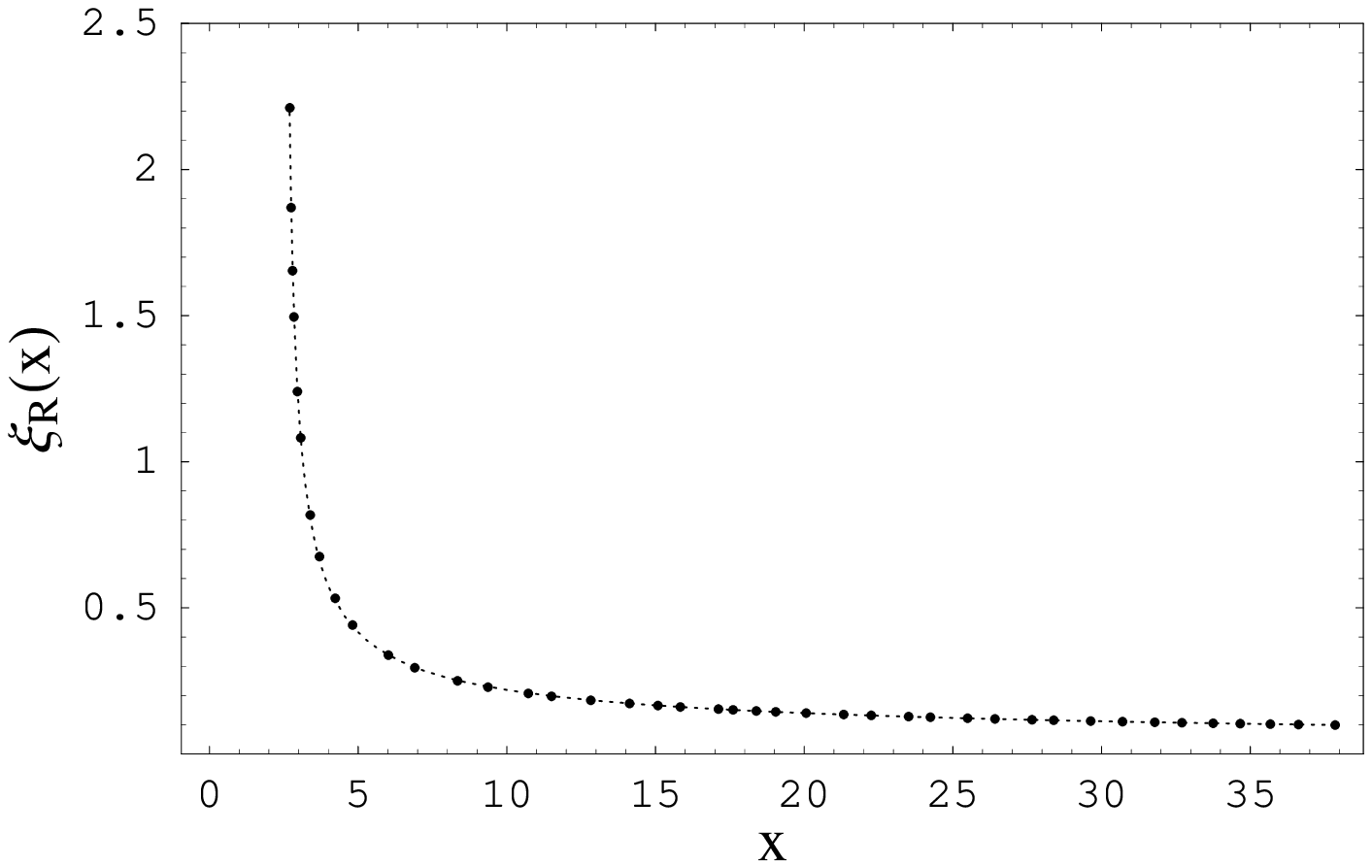}
\\
\includegraphics[clip,width=0.4\textwidth]{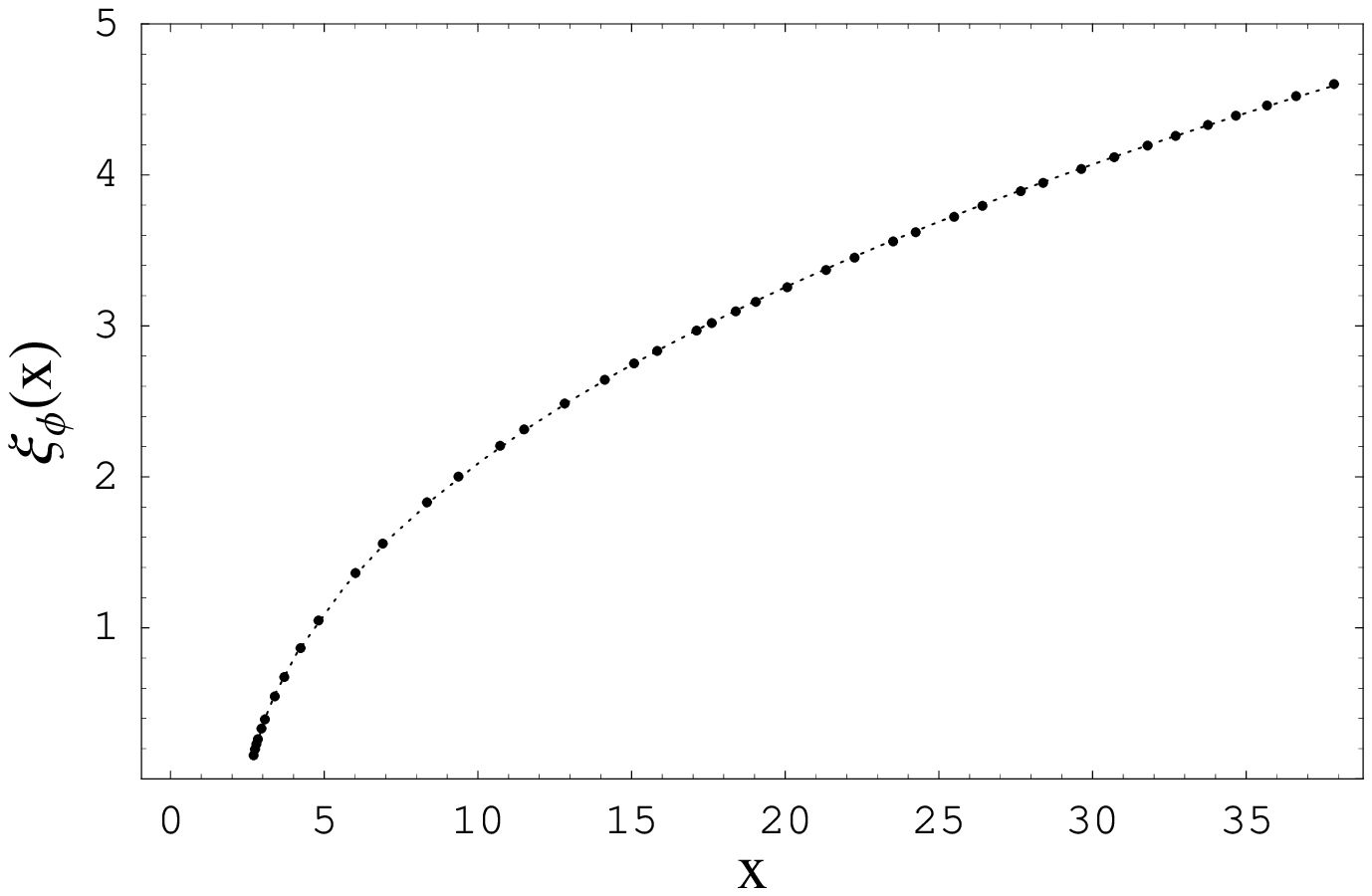}
\caption{From top to down: the functions $\xi_\omega$, $\xi_E$,
$\xi_R$, and $\xi_\phi$ defined in
Eqs.~(\ref{omega})-(\ref{phi0}). Dotted lines refer to the
approximating power-functions (\ref{xi}) (see also Table I).}
\end{center}
\end{figure}


\begin{table}[t]
\caption{Nonlinear fit of the functions $\xi$'s defined in
Eqs.~(\ref{omega})-(\ref{phi0}) using a power-function of the type
$\xi(x) = (a + b x^p)^q$. The parameters $\varepsilon_1$ and
$\varepsilon_2$ represent the maximum percentage error of the
functions $\xi$'s with respect to their numerical values in the
range $Q \in [Q_{\rm min}, 7.2 \times 10^{37} Q_{\rm cr}]$ and $Q
\in [10^{17} Q_{\rm cr}, 7.2 \times 10^{37} Q_{\rm cr}]$,
respectively.}

\vspace{0.5cm}

\begin{tabular}{lllllllll}

\hline \hline

&$$       &$~~~~~~a$  &$~~~~b$    &$~~~\,p$   &$~~~~~q$    &$~~~\varepsilon_1$ &$~~~\varepsilon_2$ \\

\hline

&$\omega$ &$~-9.567$  &$~9.686$   &$~0.381$   &$~~~~~\,1$  &$~1.32 \%$         &$~0.16 \%$ \\
&$E$      &$~-17.438$ &$~15.559$  &$~0.352$   &$~~~~~\,1$  &$~2.08 \%$         &$~0.16 \%$ \\
&$R$      &$~-7.162$  &$~4.300$   &$~0.560$   &$\:-0.712$  &$~2.76 \%$         &$~0.24 \%$ \\
&$\phi_0$ &$~-2.324$  &$~1.292$   &$~0.616$   &$~~~0.668$  &$~2.60 \%$         &$~0.23 \%$ \\

\hline \hline

\end{tabular}
\end{table}


In Fig.~5, we show the Q-ball profile for three different values
of the charge. We observe that to the charges $Q = Q_{\rm min}
\simeq 504 Q_{\rm cr}$ and $Q \simeq 10^6 Q_{\rm cr}$ there
corresponds the same value of radius, namely $R \simeq 11 m^{-1}$
(see also the third panel in Fig.~3). However, looking at the
shapes of the corresponding profiles, we see that in the first
case the profile is spreader than the second one, indicating a
larger ``wall-thickness'' of the Q-ball. Indeed, the larger the
charge, the smaller is the wall-thickness, so that, as the charge
increases the Q-ball approaches to the so-called ``thin-wall''
regime~\cite{Coleman,Dvali} (see continuous line in Fig.~5).


\begin{figure}[t]
\begin{center}
\includegraphics[clip,width=0.4\textwidth]{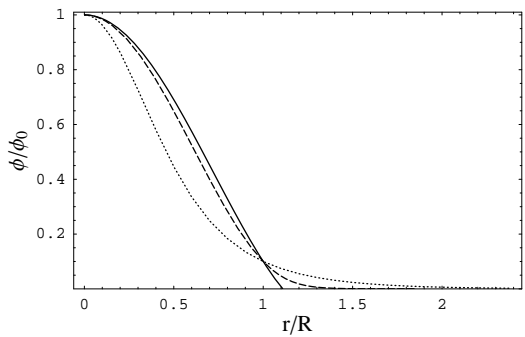}
\caption{Q-ball profile for three different values of the charge.
Dotted line: $Q = Q_{\rm min}$ ($R \simeq 11 m^{-1}$);
Dashed line $Q \simeq 10^6 Q_{\rm cr}$ ($R \simeq 11 m^{-1}$);
Continuous line $Q \simeq 7.2 \times 10^{37} Q_{\rm cr}$ ($R
\simeq 3 \times 10^8 m^{-1}$).}
\end{center}
\end{figure}


As pointed out in Ref.~\cite{de Gouvea}, for large values of the
field $\phi$, supergravity effects become important and give a
contribution to the scalar potential of the form $U_{\rm
gravity}(\phi) \simeq m_{3/2}^2 \phi^2$, approximatively. When
this contribution dominates, $U_{\rm gravity}(\phi) \gg U(\phi)$,
Q-ball properties change drastically. Indeed, a different type of
stable Q-balls are generated, the so-called ``New-type
Q-balls''~\cite{Kasuya}. The energy-charge relation is, in this
case, $E_{\rm new \, type} \sim m_{3/2} Q$ (if $Q$ is the baryon
number, being $m_p > m_{3/2}$, the Q-ball is also stable with
respect to the decay into protons). However, it is beyond the aim
of this paper to study the numerical properties of such kind of
Q-balls, and then we demand that the gauge-mediation potential
(\ref{potential}) dominates over the gravity-mediation one.
\\
Defining $\phi_{\rm eq}$ such that $U'(\phi_{\rm eq}) = U_{\rm
gravity}'(\phi_{\rm eq})$, it results $U'(\phi) \geq U_{\rm
gravity}'(\phi)$ for $\phi \leq \phi_{\rm eq}$, where a prime
indicate differentiation with respect to $\phi$. Here, following
Ref.~\cite{de Gouvea}, we have compared the derivatives of gauge-
and gravity-mediation potentials in order to determine their
relative importance, since these are the quantity entering into
the equation of motion~\eqref{eq:eqMot3}.

It is useful to introduce the ``maximum charge'' $Q_{\rm max}$
such that, if $Q \leq Q_{\rm max}$ then $\phi_0 \leq \phi_{\rm
eq}$. Since $\phi_0 = \max \phi$ for a Q-ball configuration, the
gravity effects can be neglected when $Q < Q_{\rm max}$. In
Fig.~6, we plot $Q_{\rm max}$ as a function of the gravitino mass.
It is easy to see that, neglecting logarithmic terms in the
gauge-mediation potential, one finds $U(\phi) \geq U_{\rm
gravity}(\phi)$ for $Q \leq Q_{\rm max} \simeq (m/m_{3/2})^4$
[where we used Eq.~(\ref{phi0L})]. Therefore, it is convenient to
write the maximum charge as
\begin{equation}
\label{Qmaxapprox} \frac{Q_{\rm max}}{Q_{\rm cr}} = \xi_Q \!
\left( \log_{10}\frac{m}{m_{3/2}} \right) \left( \frac{m}{m_{3/2}}
\right)^{\!4} \! ,
\end{equation}
where $\xi_Q(x)$ takes on the same form as in Eq.~(\ref{xi}). By
least-squaring the numerical data, we find $a \simeq 3.260$, $b
\simeq 5.750$, $p \simeq -1.311$, and $q \simeq 1.209$, with a
maximum percentage error on $Q_{\rm max}$ of about $0.06\%$.
Moreover, we find $Q_{\rm max} \simeq 4.9 \times 10^{28}$ and
$Q_{\rm max} \simeq 6.4 \times 10^{12}$ for $m_{3/2} = 100 \,
\keV$ and $m_{3/2} = 1\GeV$, respectively.

If $Q$ is the baryon number, it is interesting to observe that,
depending on the value of the gravitino mass, it can be $Q_{\rm
max} \lesssim 4.1 \times 10^{17} Q_{\rm cr}$, indicating that no
stable Q-ball solution exists. Numerically, we find the this
happens for gravitino masses above $m_{3/2} \gtrsim 60.8 \MeV$.


\begin{figure}[t]
\begin{center}
\includegraphics[clip,width=0.4\textwidth]{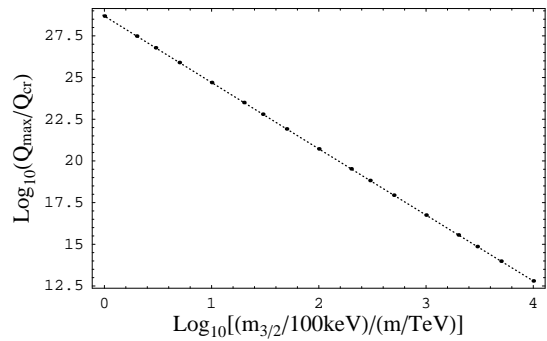}
\caption{The maximum charge $Q_{\rm max}$ as a function of the
gravitino mass. The dotted line refers to the approximating
function (\ref{Qmaxapprox}).}
\end{center}
\end{figure}



\section{IV. Conclusions}

In this paper, we have studied Q-balls-type solutions admitted in
supersymmetric particle-physics models with low-energy,
gauge-mediated supersymmetry breaking.
\\
Taking into account the exact form of the supersymmetric
potential, we have analyzed classical stability of Q-balls. We
have found, numerically, that only Q-balls with charge $Q \gtrsim
5 \times 10^2 Q_{\rm cr}$ are stable against the decay into quanta
constituting the perturbative spectrum of the theory. Here, the
``critical charge'' $Q_{\rm cr}$ is a model-dependent parameter
given in Eq.~(\ref{Qcr}).
\\
Moreover, if the conserved charge $Q$ is the baryon number,
stability with respect to the decay into protons (the lightest
baryonic particle) requires $Q \gtrsim 10^{17} Q_{\rm cr}$.

Although no analytical expressions for the quantity characterizing
Q-ball solutions can be found, we were able to approximate the
numerical results by suitable functions: we have found, indeed,
that energy and radius of Q-balls, which fully characterize their
properties from a cosmological and astrophysical viewpoint, can be
parameterized as $E \sim \xi_E Q^{3/4}$ and $R \sim \xi_R
Q^{1/4}$, where $\xi_E$ and $\xi_R$ are slowly varying functions
of the charge [see Eqs.~(\ref{energy})-(\ref{raggio}),
Eq.~(\ref{xi}), and Table I].
\\
In the (approximate) case of exactly flat potential considered in
the literature, the functions $\xi$'s are constants whose values
can differ from our results of about an order of magnitude.

For large values of the scalar condensate defining a
supersymmetric Q-ball, supergravity effects become important and
give a contribution to the scalar potential which, in turn, change
drastically the Q-ball properties. Introducing the ``maximum
charge'' $Q_{\rm max}$ such that, if $Q < Q_{\rm max}$
supergravity effects are negligible, we have found that $Q_{\rm
max}$, as a function of the gravitino mass, can be well
approximated by $Q_{\rm max} \sim \xi_Q m_{3/2}^{-4}$, with
$\xi_Q$ a slowly varying functions of $m_{3/2}$ [see
Eq.~(\ref{Qmaxapprox})].
\\
In particular, if $Q$ is the baryon number we have found that, for
gravitino masses above $m_{3/2} \gtrsim 61 \MeV$, the maximum
charge is less then the charge required for classical stability,
indicating that no stable gauge-mediation supersymmetric Q-balls
exist.



\end{document}